\documentclass[prb,twocolumn,amssymb,amsfonts,superscriptaddress,showpacs]{revtex4}
\usepackage{epsfig}
\usepackage{amsmath}
\usepackage{bm}
\usepackage{amsfonts}
\usepackage{amssymb}

\begin{document}

\title{\bf{Emission characteristics of quantum dots in planar microcavities}}

\author{G.~Ramon} \email{guy.ramon@gmail.com} \altaffiliation[Current
address: ]{Department of Physics, University of Buffalo, SUNY,
Buffalo, NY 15260-1500} \affiliation{Naval Research Laboratory,
Washington DC 20375-5320}
\author{U.~Mizrahi}
\affiliation{The Physics Department and the Solid State Institute,
Technion -- Israel Institute of Technology, Haifa 32000, Israel}
\author{N.~Akopian}
\affiliation{The Physics Department and the Solid State Institute,
Technion -- Israel Institute of Technology, Haifa 32000, Israel}
\author{S.~Braitbart}
\affiliation{The Physics Department and the Solid State Institute,
Technion -- Israel Institute of Technology, Haifa 32000, Israel}
\author{D.~Gershoni}
\affiliation{The Physics Department and the Solid State Institute,
Technion -- Israel Institute of Technology, Haifa 32000, Israel}
\author{T.~L.~Reinecke}
\affiliation{Naval Research Laboratory, Washington DC 20375-5320}
\author{B.~Gerardot}
\affiliation{Materials Department, University of California Santa
Barbara, CA, 93106}
\author{P.~M.~Petroff}
\affiliation{Materials Department, University of California Santa
Barbara, CA, 93106}

\begin{abstract}

The emission properties of single quantum dots in planar
microcavities are studied experimentally and theoretically.
Fivefold Enhanced spontaneous emission outside the microcavity is
found for dots in resonance with the cavity mode, relative to
detuned dots, while their radiative lifetime is only marginally
decreased. Using high power excitation we obtain the in-plane
cavity dispersion. Near field images of the emission show spatial
distributions of several microns for resonant dots, which decrease
in size with the detuning from resonance. These experimental
findings are explained using a simple and intuitive model.

\end{abstract}


\maketitle

\section{Introduction}

The potential applications of semiconductor quantum dots (QDs) as
quantum light emitters have generated considerable research
efforts in recent years.\cite{Michler-Santori-Regelman} Single
photon sources using quantum dots are an important ingredient in
quantum information applications such as quantum cryptography and
teleportation,\cite{Kiraz} and may be further employed to
implement efficient linear optics quantum computation \cite{Knill}
with the attainment of on-demand indistinguishable single-photon
pulses.\cite{Santori} Single QDs can be excited either with
continuous wave (cw), where they emit antibunched photons obeying
a sub-Poissonian statistics or with pulsed light, where a single
photon is emitted per pulse.\cite{Michler-Santori-Regelman} It is
the radiative recombination of electron-hole (e-h) pairs (or
excitons) confined in single QDs that gives rise to the emission
of single photons.

QD based devices have several advantages as single photon
emitters. These include relatively large oscillator strengths,
narrow spectral linewidths, and compatibility with mature
semiconductor technologies. A major hindrance in their usefulness,
though, is the low extraction efficiency of the emitted photons
due to the high refraction index of the host semiconductor.
Typically, only about two percent of the photons will be emitted
from a GaAs based device, while the rest will be lost due to total
internal reflection. The general approach to overcome this problem
has been to place the QDs at the antinode of a microcavity whose
dimensions are comparable to the wavelength of the emitted
photons.\cite{Vahala} In these cavities the number of allowed
optical modes is reduced and at the same time,  the in-cavity
intensity of the allowed modes increases. As a result, the
spontaneous recombination rate of excitons within these resonating
modes is increased, a phenomenon known as the Purcell
effect.\cite{Gerard,Pelton}

Several approaches to realize 3D photon confinement have been
studied, including whispering-gallery modes in microdisks,
\cite{Garyal} defect modes in 2D photonic crystals,\cite{Yoshie}
and lateral patterning of planar dielectric cavities using
electron-beam lithography and etching.\cite{Pelton,Reithmaier} The
latter has proved particularly useful, demonstrating efficient
photon emission in the weak coupling regime,\cite{Pelton} and more
recently strong coupling, paving the pathway for solid state
realization of coherent control schemes.\cite{Reithmaier} In the
strong coupling regime, the interaction between the cavity mode
and the emitter is larger than their combined decay rates, and the
irreversible spontaneous emission from the QD exciton is replaced
by a coherent exchange of energy between the exciton and the
cavity mode. Among the challenges in the implementation of strong
coupling in structured cavities, one notes the reduction in the
cavity's Q-factor due to the lateral patterning, and the lack of
control in the in-plane positioning of the QD  with respect to the
lateral microstructure, resulting in a reduction in the oscillator
strength of the confined exciton.\cite{Reithmaier,Badolato}

In view of the technological difficulties associated with systems
of QDs in 3D cavities, it seems important to study a system of QDs
embedded in a planar microcavity, which is easier to fabricate.
These cavities consist of distributed Bragg reflectors (DBRs),
which are typically stacks of alternating quarter-wavelength-thick
layers of GaAs and AlAs, separated by a spacer layer of GaAs.
Planar cavities can attain high Q factors due to the excellent
control in their fabrication and embedding the QDs at the field's
antinode is straightforward. These structures support an in-plane
continuum of modes, and the interaction with QD excitons is
therefore not expected to be in the strong coupling
regime.\cite{Andreani} Nevertheless, the spontaneous emission and
the light extraction efficiency can still be enhanced
considerably. \cite{Bjork91} For example, a relatively weak planar
cavity has been utilized recently to produce a ten-fold increase
in the efficiency of light collection of GaAs based LED
structures.\cite{Bennett}

Unlike systems of quantum wells embedded in planar cavities, where
polariton effects prevail and have been extensively studied (see
e.g., \cite{Weis,Savona1,Savona2,Citrin}), QDs in planar cavities
have received little attention.\cite{Sug2,Andrews} Among the few
experimental studies of this system, angle-resolved
photoluminescence spectroscopy was used to detect TE-TM mode
splitting (not to be confused with Rabi splitting, which is
discernable only in the strong coupling regime).\cite{Hu}

In the present work we study experimentally and theoretically
single QD emitters in a planar microcavity. Using a partially
covered layer of InGaAs self assembled QDs placed at the center of
a $\lambda$ microcavity, we have measured various characteristics
of the spontaneous emission dynamics of the system such as spatial
and angular distributions of the radiation field and its temporal
decay. Our results are compared with a relatively simple
theoretical analysis, which follows the treatment of Sugawara
\cite{Sug2} and gives a simple and intuitive picture of the
spontaneous emission properties of recombining excitons confined
in QDs embedded in a planar microcavity. This approach accounts
reasonably well for our experimental findings, including emission
rates, radiative lifetimes, and angular and spatial distributions
of the emission. In all of these measurements, a crucial parameter
is found to be the detuning of the emitter energy from that of the
cavity resonant mode. Since we do not have an exact knowledge of
the QDs size, shape and composition, detuning from resonance is
modeled in our approach by either varying the QD size or by
varying its material bandgap, while keeping the optical properties
of the  microcavity fixed.

The paper is organized as follows. In section \ref{exp} we provide
the details of our experimental system. Section \ref{theory} gives
the theoretical model including the exciton wavefunctions and
their interaction with the cavity field. In section \ref{results}
we give the experimental measurements and compare them with model
calculations. A summary is given in section \ref{summary}, and
some details pertaining to the calculation of the excitonic
wavefunction are given in the appendix.

\section{Experimental setup}
\label{exp}

The samples were grown by molecular beam epitaxy on a (100)
oriented GaAs substrate. One layer of strain-induced InGaAs QDs
was deposited in the center of a one wavelength GaAs spacer layer.
The height and composition of the QDs were controlled by partially covering the
InAs QDs by $30\AA$ thick layer of GaAs and by subsequent two
minutes growth interruption\cite{Garcia} to allow diffusion of In (Ga) atoms
from (into) the strained islands. The growth resulted in ${\rm
In}_{{\rm x}}{\rm Ga}_{1-{\rm x}}{\rm As}$ QDs whose estimated
radius and composition are $150\AA \lesssim R_{\rm QD} \lesssim
250\AA$ and 0.47$\lesssim {\rm x} \lesssim 0.53$, respectively,
and height of 30\AA. The GaAs spacer layer was grown to a width
whose resonance mode matches the QDs emission due to ground state
e-h pair recombinations (1 $\lambda$ cavity). The microcavity was
formed by 25 and 11 period stacks of alternating quarter
wavelength layers of AlAs and GaAs below and above the spacer
layer, respectively. The samples were not patterned or processed
laterally to prevent obscuring the emission and its polarization.

For the optical measurements we used a diffraction limited low
temperature confocal optical microscope \cite{Dekel98, Dekel00}.
The sample was cooled by a copper braid attached to the cold
finger of a He-flow cryostat. The sample mount was accurately
manipulated in three directions using computer-controlled motors.
A X60 in-situ microscope objective was used in order to focus cw
or pulsed laser light at normal incidence on the sample. The
emitted light was collected by the same microscope objective. The
collected light was spatially filtered, dispersed by a 0.22 m
monochromator and detected by a nitrogen-cooled CCD array
detector. The system provides diffraction-limited spatial
resolution, both in the excitation and the detection channels. For
the time-resolved spectroscopy, the dispersed light from the
monochromator was focused onto a small, thermoelectrically cooled,
single channel avalanche silicon photodiode. The signal from the
photodiode was analyzed using conventional photon counting
electronics. The photodiode dark count rate was ~100 ${\rm
sec}^{-1}$, and the system temporal resolution was approximately
250 psec.

\section{Theoretical framework}
\label{theory}

The following approach assumes a single 1s exciton confined in a
QD interacting with a discrete cavity mode. No charged exciton or
biexciton states were considered, although their presence was
confirmed in the measurements (see section \ref{resultsA}). Our
main interest in the current study is in the cavity effects on the
excitons emission dynamics. The assumption of discrete cavity
modes is justified for high finesse cavities. (we have verified
that only the basic $\lambda$ mode contributes appreciably to the
interaction with the exciton). Although more complete description
of the electromagnetic density of states in the cavity is
available \cite{Savona2}, the main features of the exciton-photon
coupling are well demonstrated within the framework of our
simplified model.

\subsection{Model wavefunction for quantum dot excitons}
\label{ex-wf} 

We start by solving for the excitonic wave function in a quantum
dot, considering finite potential barrier in the growth ($z$)
direction and parabolic potential in the lateral direction.

The effective mass Hamiltonian is given by
\begin{eqnarray}
H&=& -\frac{\hbar^2}{2m_e^{\|}} \left[ \nabla^2_{\rho_e}+\sigma
\nabla^2_{\rho_h}+\sigma_{ez}\partial^2_{z_e} +\sigma_{hz}
\partial^2_{z_h} \right] - \nonumber \\ && \frac{e^2}
{\epsilon \sqrt{(\mbox{\boldmath $\rho$}_e-\mbox{\boldmath
$\rho$}_h)^2+ (z_e-z_h)^2}}+ \nonumber \\
&&  V_e(\mbox{\boldmath $\rho$}_e,z_e)+V_h(\mbox{\boldmath
$\rho$}_h,z_h), \label{H1}
\end{eqnarray}
where ${\bf r}_i=(\mbox{\boldmath $\rho$}_i,z_i)$ are the electron
or hole in-plane and $z$ coordinates, and
$\sigma=m_e^{\|}/m_h^{\|}$, $\sigma_{ez}=m_e^{\|}/m_e^{z}$,
$\sigma_{hz}=m_e^{\|}/m_h^{z}$ are the appropriate effective mass
ratios in the plane and $z$ directions, and $\epsilon$ is the
background dielectric constant screening the Coulomb interaction.
In Eq.~(\ref{H1}) we have neglected the differences in the
conduction- and valence-band masses and in the dielectric constant
value of the two semiconductors that comprise the QD and its host.
In the case of strong $z$ direction confinement, the electron and
hole potentials can be approximated as
\begin{equation}
V_i({\mbox{\boldmath $\rho$}}_i,z_i) \approx V_i(\mbox{\boldmath
$\rho$}_i)+V_i(z_i),\ \ \ i=e,h
\end{equation}
thus decoupling the problem into lateral and $z$ parts, and
enabling us to write the envelope wave function for the exciton in
the form
\begin{equation}
\Phi_{\rm x}({\bf r}_e, {\bf r}_h)=\Psi_{\rm x}(\mbox{\boldmath
$\rho$}_e,\mbox{\boldmath $\rho$}_h) \chi_e(z_e) \chi_h(z_h)
\end{equation}
(for infinite $z$ direction potential, this approximation is
exact). Transforming to center of mass (CM) and relative in-plane coordinates:
\begin{equation}
{\bf R}=\frac{\sigma \mbox{\boldmath $\rho$}_e+\mbox{\boldmath
$\rho$}_h}{1+\sigma} \ \ ; \ \ {\bf r}=\mbox{\boldmath
$\rho$}_e-\mbox{\boldmath $\rho$}_h,
\end{equation}
we separate the CM-motion and relative-motion wave functions,
$\Psi_{\rm x}({\bf R}, {\bf r})=\psi({\bf R}) \phi({\bf r})$, by
Choosing parabolic potentials for the lateral confinement of both
electron and hole\cite{Que,CM}
\begin{equation}
V_i(\mbox{\boldmath
$\rho$}_i)=\frac{1}{2}m_i^{\|}\omega^2\rho_i^2, \ \ i=e,h.
\label{harmonic}
\end{equation}
Employing natural units of length and energy, namely, bulk
effective Bohr radius ($a_B=\hbar^2 \epsilon/m_e^{\|} e^2$) and
Rydberg (${\rm Ry}=\hbar^2/2 m_e^{\|} a_B^2$), the Hamiltonian in
Eq.~(\ref{H1}) takes the form
\begin{eqnarray}
H & \!\!= \!\! & H_R+H_r, \nonumber \\
H_R& \!\!=\!\!&-\frac{\sigma}{1+\sigma} \nabla^2_R +\frac{4 \sigma
R^2}{\xi^4(1+\sigma)} \label{H2} \\
H_r&\!\!=\!\!& -\left[(1+\sigma) \nabla^2_r+\sigma_{ez}
\partial^2_{z_e}+\sigma_{hz}
\partial^2_{z_h} \right]- \nonumber \\
&& \frac{2}{\sqrt{r^2+(z_e-z_h)^2}}+ \frac{4 \sigma^2
r^2}{\xi^4(1+\sigma)^3} +V_e(z_e)+V_h(z_h) \nonumber
\end{eqnarray}
where $\xi$ is a dimensionless lateral localization parameter
related to the confining potential by
\begin{equation}
\xi=\frac{1}{a_B} \sqrt{\frac{2\hbar}{M_{\rm x} \omega}}
\end{equation}
and $M_{\rm x}=m_e^\|+m_h^\|$ is the in-plane CM exciton mass.

We identify the in-plane CM motion part with a 2D harmonic
oscillator Hamiltonian, which is solved analytically resulting in
wave functions $\psi({\bf R})$ which are given by the associated
Laguerre polynomials \cite{Sug1}. In what follows, we will be
interested only in the CM-motion ground state which is a 2D
Gaussian function. By equating the areal size of the CM wave
functions with the physical QD radius: $|\int \psi({\bf R}) d^2
R|^2 =2\pi \xi^2= \pi R_D^2$, we relate the potentials
(\ref{harmonic}) with the QD size.

For the in-plane relative coordinates we employ a trial wave
function \cite{Sug1}
\begin{equation}
\phi(r)={\cal N}_r {\rm e}^{-(r/\eta)^\alpha} \label{phi}
\end{equation}
where ${\cal N}_r$ is a normalization constant and $\eta, \alpha$
are variational parameters. $\alpha$ has values between 1 (no
lateral confinement - very large QDs) and 2 (strong lateral
confinement - very small QDs). Considering finite height
potentials in the $z$ direction, $V_i(z_i)=V_i^z
\Theta(|z_i|-L/2)$ where $i=e,h$ and $L$ is the QD height, we have
for the $z$ wave functions \cite{GreBaj,GruBim}
\begin{equation}
\chi_i(z_i)= {\cal N}_{z_i} \times \left\{
\begin{array}{ll}
\cos (k_i z_i) \ & \left|z_i\right| \leq L/2 \\
\frac{k_i}{\sqrt{k_i^2+\kappa_i^2}} {\rm e}^{-\kappa_i(|z_i|-L/2)}
\ & \left| z_i \right|
>L/2
\end{array}
\right. \label{chi}
\end{equation}
where $i=e,h$, ${\cal N}_{z_i}=\left( L/2+1/\kappa_i
\right)^{-1/2}$ is a normalization constant,
\begin{equation}
k_i=\sqrt{\frac{2 m_i^z}{\hbar^2} E_{i0}^z} \ \ , \ \
\kappa_i=\sqrt{\frac{2 m_i^z}{\hbar^2}(V_i^z-E_{i0}^z)},
\end{equation}
and $E_{i0}^z$ is determined by the ground state quantization
condition:
\begin{equation}
\sqrt{\frac{E_{i0}^z}{V_i^z}}=\cos \left(\frac{L}{2}\sqrt{\frac{2
m_i^z}{\hbar^2}E_{i0}^z}\right).
\end{equation}

The variational parameters in Eq.~(\ref{phi}) are calculated by
maximizing the magnitude of the ground state exciton binding
energy $E_B=E_{e0}^z+E_{h0}^z+E_{\rm 1s}^{\rm CM}-E_{\rm r}$,
where $E_{\rm 1s}^{\rm CM}=\hbar \omega=\frac{4 \sigma}{\xi^2
(1+\sigma)}$ is the ground state energy of the CM motion and
$E_{\rm r}$ is the expectation value of the relative motion part
of the Hamiltonian in (\ref{H2}) (see Appendix).

\begin{figure}[htbp]
\epsfxsize=1.1\columnwidth
\vspace*{0.1cm} \centerline{\epsffile{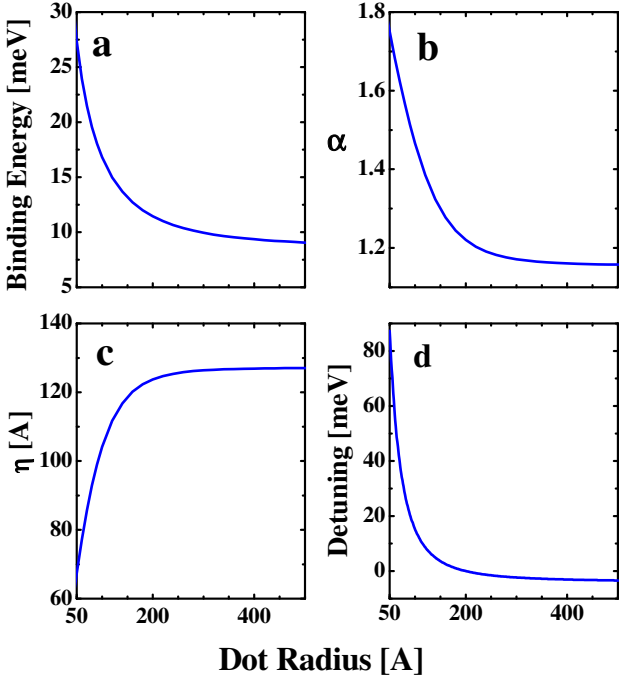}}
\vspace{-1.8cm} \caption{(a) Exciton binding energy, (b)
Variational parameter $\alpha$,  (c) Variational parameter $\eta$,
and (d) Detuning of the exciton line with respect to the cavity
mode, as functions of the QD radius (resonance was assumed for
$R_D=200 \AA$).} \label{binding}
\end{figure}

The InGaAs self-assembled QDs that we are considering have an
estimated height of $L=30$\AA. The conduction- and valence-band
in-plane and normal to the plane masses were calculated for this
quantum well width, assuming $50\%$ InAs contents, using an eight
band Kane model, and taking into account both non-parabolicity and
lattice-mismatch strain effects.\cite{GershoniJQE} We find for the
electron: $m_e^\|=0.0630$, $m_e^z=0.0897m$, and for the heavy
hole: $m_h^\|=0.1573m$, $m_h^z=0.3406m$. For the potential
barriers we find: $V_e^z=441.9$meV and $V_h^z=110.6$meV. Using
these values we minimize Eq.~(\ref{Eexp}) with respect to the two
variational parameters $\alpha$ and $\eta$, and find the binding
energy of the ground state heavy-hole exciton as a function of the
QD radius shown in figure \ref{binding}.

\subsection{Exciton-photon interaction in a planar microcavity}

In the following we assume weak coupling between the QD exciton
and the cavity modes. This is justified a-posteriori by the
calculated spontaneous emission linewidths, which are always much
smaller than that of the cavity. The cavity photon escape rate was
deduced from the measured reflectivity of the cavity. We have put
the cavity structure into a linear dispersion model and calculated
it's reflectivity using transfer matrix formalism. The resulting
cavity mode linewidth was close to the measured value.

The interaction between an exciton and the electromagnetic field
is given by
\begin{equation}
H_{\rm int} = -\frac{e}{mc} \sum_{{\bf k},\sigma} {\bf p} \cdot
{\bf A}_{\bf k}^\sigma
\end{equation}
where ${\bf A}_{\bf k}^\sigma$ is the vector potential associated
with the $k$th mode with polarization $\sigma$ and we have
neglected the quadratic term in ${\bf A}$. In a planar cavity,
with $z$ in the growth direction, the Maxwell equations with the
appropriate boundary conditions yield two solutions for the
electromagnetic field.\cite{Jackson} These correspond to the
electric vector being normal to (TE) or in (TM) the plane defined
by the wave vector ${\bf k}=({\bf k}_\|,k_z)$ and $z$, and are
given (using the Coulomb gauge) by:
\begin{subequations}
\begin{eqnarray}
{\bf A}_{\bf k}^{{\rm TE}}&\!\!=\!\!&-{\rm i} \sqrt{\frac{2 \pi
\hbar v}{k V_c}} \cos (k_z z){\bf e}_\| \left[{\rm e}^{\rm i {\bf
k}_\| \cdot \mbox{\boldmath $\rho$}} \hat{a}_{\bf k}+{\rm e}^{-\rm
i {\bf k}_\| \cdot \mbox{\boldmath
$\rho$}} \hat{a}^\dag_{\bf k} \right] \\
{\bf A}_{\bf k}^{{\rm TM}}&\!\!=\!\!& \sqrt{\frac{2 \pi \hbar v}{k
V_c}} \left[ -{\rm i} \frac{k_z}{k} \cos (k_z z){\bf e}_\|
+\frac{k_\|}{k} \sin ( k_z z){\bf e}_\perp \right] \times \nonumber \\
&& \left[{\rm e}^{\rm i {\bf k}_\| \cdot \mbox{\boldmath $\rho$}}
\hat{a}_{\bf k}+{\rm e}^{-\rm i {\bf k}_\| \cdot \mbox{\boldmath
$\rho$}} \hat{a}^\dag_{\bf k} \right]
\end{eqnarray}
\label{A}
\end{subequations}
where $v=c/n_{\rm eff}$ is the velocity of light in the cavity
medium, $V_c$ is the cavity quantization volume, and $\hat{a}_{\bf
k}^\dag$ ($\hat{a}_{\bf k}$) is the creation (annihilation)
operator of the ${\bf k}$ field mode. In Eqs.~(\ref{A}) $k_z$
satisfies the resonance condition
\begin{equation}
k_z=\frac{2m \pi}{L_c},
\end{equation}
where $m$ takes any integer value and $L_c$ is the cavity width.
Here we have considered only even modes which are the only ones
which couple with the exciton ground state.\cite{Savona1} Assuming
a system of a single exciton and a single photon, appropriate for
low exciton densities, the time-dependent state of the system is
given by
\begin{equation}
|\Psi(t) \rangle=b(t){\rm e}^{-{\rm i} \omega_{\rm x} t}
|\Phi_{\rm x},0 \rangle+\sum_{{\bf k},\sigma} c_{{\bf
k},\sigma}(t) {\rm e}^{-{\rm i} \omega_k t} |\Phi_{\rm g},1_{{\bf
k},\sigma} \rangle
\end{equation}
where $\omega_{\rm x}$ ($\omega_k$) is the exciton
(electromagnetic field) resonant frequency, and $\sigma$ runs over
the two light polarizations. Taking the system to be initially
with one exciton and the radiation field in the vacuum state
[$b(0)=1$, $c_{{\bf k},\sigma}(0)=0$] we write the equations of
motion for $b(t)$ and $c_{{\bf k},\sigma}(t)$, using the rotating
wave approximation:
\begin{subequations}
\begin{eqnarray}
\dot{b}(t)&\!\!=\!\!&-{\rm i}\sum_{{\bf k},\sigma}g_{\bf k,
\sigma}{\rm e}^{{\rm i}(\omega_{\rm x}-\omega_k)t}c_{{\bf
k},\sigma}(t)-\frac{\gamma_{\rm x}}{2} b(t) \label{a} \\
\dot{c}_{{\bf k},\sigma}(t)&\!\!=\!\!&-{\rm i}g^*_{\bf k,
\sigma}{\rm e}^{-{\rm i}(\omega_{\rm x}-\omega_k)t}b(t)-
\frac{\gamma_{\rm c}}{2} c_{{\bf k},\sigma}(t). \label{b}
\end{eqnarray}
\label{eq_motion}
\end{subequations}
In Eqs.~(\ref{eq_motion}) $\gamma_{\rm x}$ is the exciton
broadening due to all processes other than spontaneous emission
(phonon scattering, nonradiative recombination), and $\gamma_{\rm
c}$ is the decay rate of the cavity photon mode due to mirror
losses, which is inversely proportional to the cavity quality
factor. The exciton-photon coupling constant appearing in
Eqs.~(\ref{eq_motion}) is given by
\begin{equation}
\hbar g_{\bf k, \sigma}=-\langle \Phi_{\rm x},0|\frac{e}{mc} {\bf
p} \cdot {\bf A}_{\bf k}^\sigma | \Phi_{\rm g},1_{{\bf k},\sigma}
\rangle.
\end{equation}
The coupling constant is related to the total oscillator strength
per unit area $f_{\bf k}$ through
\begin{equation}
\sum_{\sigma={\rm TE},{\rm TM}} |g_{\bf k, \sigma}|^2 = \frac{2
\pi e^2 \omega_{\rm x}}{n_{\rm eff} mckL_c} f_{\bf k}.
\end{equation}
Using the dipole approximation (${\bf k}_e \approx -{\bf k}_h$),
considering both polarizations and neglecting inter-subband
mixing, $f_{\bf k}$ is calculated to be
\begin{equation}
f_{\bf k}=|\tilde{\psi}(k_\|)|^2 \left[ \left( 1+\frac{k_z^2}{k^2}
\right) \cos^2 (k_z z) f_\|+\frac{k_\|^2}{k^2} \sin^2 (k_z z)
f_\perp \right] \label{fk}
\end{equation}
where
\begin{subequations}
\begin{eqnarray}
f_\| &=& \frac{|\phi (0)|^2}{m \hbar \omega_{\rm x}} | P_{\rm cv}
F_{{\bf e}_\|}|^2 | \langle \chi_e | \chi_h
\rangle |^2 \\
f_\perp &=& \frac{|\phi (0)|^2}{m \hbar \omega_{\rm x}} | P_{\rm
cv} F_{{\bf e}_\perp} |^2|\langle \chi_e | \chi_h \rangle |^2.
\end{eqnarray}
\label{fpp}
\end{subequations}
In Eqs.~(\ref{fk}-\ref{fpp}) $\tilde{\psi}(k_\|)$ is the in-plane
Fourier transformed exciton CM wave function, $\phi (0)$ is the
in-plane exciton relative motion wave function at zero, $P_{\rm
cv}$ is the bulk transition matrix element, and $F_{{\bf e}_\|}$
($F_{{\bf e}_\perp}$) is the parallel (perpendicular) polarization
factor in quantum wells.\cite{Sug1}

Integrating Eq.~(\ref{b}) and substituting the resulting $c_{{\bf
k},\sigma}(t)$ in Eq.~(\ref{a}) we have the integro-differential
equation
\begin{eqnarray}
\dot{b}(t)&=&-\sum_{{\bf k},\sigma} |g_{{\bf k},\sigma}|^2
\int_0^t d\tau b(\tau){\rm e}^{[-{\rm i}(\omega_{\bf
k}-\omega_{\rm
x})-\gamma_{\rm c}/2] (t-\tau)}- \nonumber \\
&& \frac{\gamma_{\rm x}}{2}b(t). \label{intdif}
\end{eqnarray}
In the weak coupling regime, which is clearly our case,
$\gamma_{\rm c} \gg g$ and one can solve Eq.~(\ref{intdif}) to a
good approximation by taking $b(t)$ out of the integral. For $t
\gg \gamma_{\rm c}^{-1}$ the result is
\begin{equation}
\dot{b}(t)=-\frac{\gamma_{\rm x}+\gamma_{\rm SE}}{2} b(t)
\label{adot}
\end{equation}
where
\begin{eqnarray}
\gamma_{\rm SE}&=&2\pi \sum_{{\bf k},\sigma}
\mbox{\raisebox{1.5ex}{$\scriptstyle{\prime}$}} |g_{{\bf
k},\sigma}|^2 {\cal L}(\omega_{\bf k}-\omega_{\rm x},\gamma_{\rm
c}) \nonumber \\
&=&\frac{4\pi^2 e^2 \omega_{\rm x}}{n_{\rm eff}^2 mL_c} \sum_{\bf
k} \mbox{\raisebox{1.5ex}{$\scriptstyle{\prime}$}} \frac{f_{\bf
k}}{\omega_{\bf k}} {\cal L}(\omega_{\bf k}-\omega_{\rm
x},\gamma_{\rm c}), \label{SE}
\end{eqnarray}
and
\begin{equation}
{\cal L}(\omega_{\bf k}-\omega_{\rm x},\gamma_{\rm c})=
\frac{1}{2\pi} \frac{\gamma_{\rm c}}{(\omega_{\bf k}-\omega_{\rm
x})^2+(\gamma_{\rm c}/2)^2}
\end{equation}
is the normalized Lorentzian cavity-mode broadening. The prime
over the sum in Eq.~(\ref{SE}) indicates taking the lowest cavity
mode, $k_z=2\pi/L_c$, and summing over the in-plane photon wave
vectors, ${\bf k}_\|\leqslant q_{\rm x}$, where $q_{\rm
x}=\omega_{\rm x}/v$ is the maximum exciton in-plane wave vector
that can still couple to the cavity mode resulting in radiative
recombination of the exciton. The solution of Eq.~(\ref{adot})
gives an exponential irreversible spontaneous emission of the
exciton due to its coupling with the cavity field modes,
representing Fermi's golden rule. As long as the calculated
$\gamma_{\rm SE}$ is much smaller than the cavity broadening
$\gamma_{\rm c}$, the weak coupling approximation holds;
otherwise, one must solve the general equation (\ref{intdif}).

\section{Results}
\label{results}

We are interested in particular in the influence of the cavity on
the exciton's spontaneous emission, radiative lifetime, and
emission distribution.

\subsection{Exciton Spontaneous Emission}
\label{resultsA}

\begin{figure}[!htbp]
\epsfxsize=1.1\columnwidth
\vspace*{1.4cm} \centerline{\hspace{-0.4cm} \epsffile{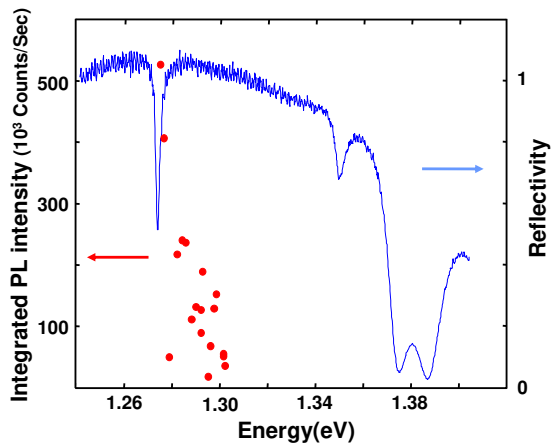}}
\vspace*{-2.8cm} \caption{(color online) Measured reflectivity
(blue line) and integrated photoluminescence intensity (red
points) of QDs at the antinode of a planar $\lambda$ cavity.}
\label{dudi1}
\end{figure}

Figure \ref{dudi1} shows the measured reflectivity of the cavity
and the integrated photoluminescence intensity, using relatively
high power, constant excitation. Each point in the figure results
from emission from a distinct QD. The numerical aperture of our
microscope objective in the PL measurements was 0.8, resulting in
a collection angle of $\sim 13.2^\circ$ inside the sample. The
figure shows a sharp dip in the reflectivity corresponding to the
resonant cavity mode. The FWHM is 1.94meV giving a Q-factor of
670. The emission intensity shows a pronounced enhancement for QDs
in resonance with the cavity mode. A rapid decrease in the
emission intensity to a value roughly five times smaller is found
for QDs detuned by $ 20-25 {\rm meV}$.

In order to discuss these data, we relate the detuned exciton line
to the $\lambda$ cavity width by\cite{Sug2}
\begin{equation}
L_c=\frac{\lambda_{\rm x}}{n_{\rm eff}} \zeta =\frac{2 \pi}{q_{\rm
x}} \zeta,
\end{equation}
where $\zeta$ corresponds to the detuning of the exciton line from
resonance. We assume the cavity resonant mode matches the emission
from the heavy hole 1s exciton of a QD with a radius $R_{\rm
D}=200{\rm \AA}$, for which $\zeta =1$. The corresponding
resonance energy is calculated using Eq.~(\ref{Eexp}) to be
$E^{\rm res}_{\rm x}=E_c=1.2756$eV. In order to model
microscopically the emission from quantum dots that are detuned
from this resonance, we represent the detuning by either changing
the QD size or by changing the QD material bandgap (via the dot
composition). The dependence of the detuning on the QD radius is
given in Fig.~\ref{binding}d. Considering only the basic cavity
mode expressed as $k_z=q_{\rm x}/\zeta$, Eqs.~(\ref{SE}) and
(\ref{fk}) are used to give
\begin{eqnarray}
\gamma_{\rm SE}&\!\!=\!\!& \frac{\pi e^2 \omega_{\rm x}^2 R_{\rm
D}^2 f_\|}{mc^2 \zeta} \int_0^{q_{\rm x}} dq q {\rm e}^{-R_{\rm
D}^2 q^2/4} \frac{q^2+2 q^2_{\rm x}/\zeta^2}{(q^2+q_{\rm
x}^2/\zeta^2)^{3/2}} \times \nonumber \\
&& {\cal L} \left( v\sqrt{q^2+q_{\rm x}^2/\zeta^2}-\omega_{\rm x}
, \gamma_c \right). \label{SE2}
\end{eqnarray}
In deriving Eq.~(\ref{SE2}) we assumed that the QD resides in the
center of the cavity, where the cavity mode takes its maximum
value, and considered only the heavy-hole exciton whose
perpendicular transition matrix element vanishes, $f_\perp=0$. To
account for the experimental collection angle, the upper limit of
the integration in Eq.~(\ref{SE2}) is replaced with $q_0=q_{\rm x}
\sin \theta$, where $\theta$ is measured from the $z$-axis.

In the calculation we used $n_{\rm eff}=3.5$, $\epsilon=13.9$
resulting in QW exciton Bohr radius of $a_B=116.7$\AA. The free
(QW) exciton oscillator strength per unit area was calculated for
a $30{\rm \AA}$ ${\rm In}_{0.53}{\rm Ga}_{0.47}{\rm As}$ QW and
was found to be $f_\|=7.1 \cdot
10^{-5}$\AA$^{-2}$.\cite{Sug1,Andreani90} This value increases as
the QD size reduces, due to the shrinkage of the relative motion
wave function (see Eqs.~\ref{fpp}). We took an additional factor
of two for the oscillator strength to account for the two spin
configurations.

\begin{figure}[!htbp]
\epsfxsize=0.85\columnwidth
\vspace*{1.8cm} \centerline{\hspace{-0.2 cm} \epsffile{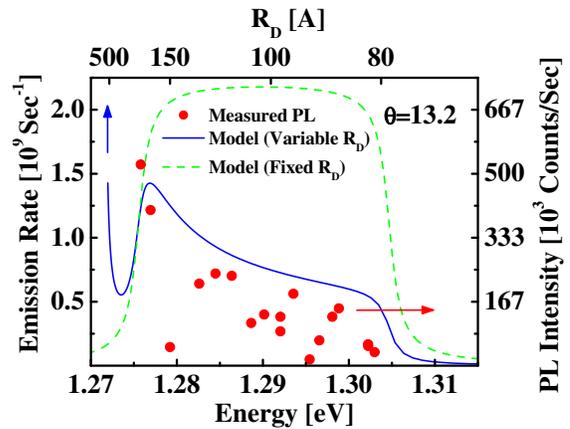}}
\vspace{-1.9cm} \caption{(color online) Emission rates into a
collection angle $13.2^\circ$ vs. exciton energy. Full (red)
circles are the measured values given in counts per second (right
axis). Solid (blue) line shows the calculation where detuning is
obtained by varying the QD radius (given in the top axis). Dashed
(green) line shows the calculation with QD size fixed to $R_{\rm
D}=200\AA$ and the detuning obtained by varying the QD material
bandgap. A scaling factor of 3000 was taken between the left and
right axes (see text).} \label{rate}
\end{figure}

In Figure \ref{rate} the calculated exciton spontaneous emission
rates are shown together with the measured PL values as functions
of exciton energy. The solid (blue) line corresponds to a
calculation where detuning of the exciton energy was obtained by
varying the QD radius whereas for the dashed (green) line the QD
size was fixed to $R_{\rm D}=200\AA$ and its bandgap was varied.
In order to put the measured and calculated values at the same
figure, a scaling factor of 3000 was taken between the right
(photon counts) and left (emission rate) axes. This factor
corresponds to the photon extraction efficiency in the experiment,
which was estimated from pulsed excitation
measurements.~\cite{Dekel00}

The calculated results obtained by varying the dot size agree
qualitatively with the experimental data, including the rapid fall
off in the PL intensity away from resonance. The range of $90 \AA
\lesssim R_D\lesssim 300 \AA$ that accounts for a detuning range
of $\sim 20 {\rm meV}$ is consistent with variations in typical
self-assembled InGaAs QDs. The two model calculations in
Fig.~\ref{rate} show marked differences. In particular, the
emission enhancement shown for negative detuning in the case of
variable QD size is missing in the fixed QD size calculation. The
latter case shows a rapid inhibition of the emission as the QD
becomes negatively detuned out of the cavity mode linewidth, which
is clearly a cavity effect. In the case of variable $R_{\rm D}$
the cavity effect competes with the localization of the exciton CM
wave function in ${\bf k}_\|$ space, as the QD size increases
[Eq.~(\ref{SE2})]. This effect is consistent with the strong
increase in QD exciton lifetime as compared with QW excitons. The
competition between the cavity and localization effects is shown
in the figure by the dip in the calculated emission rate (solid
line) for $R_{\rm D} \sim 300{\rm \AA}$ (negative detuning). For
yet larger QDs the localization effect takes over and the emission
rate rapidly increases (the figure includes QDs with $R_{\rm
D}\leq600{\rm \AA}$). Such large QDs are less likely to form in
our case and this increase is therefore not observed.

\begin{figure}[!htbp]
\epsfxsize=1.0\columnwidth
\vspace*{1.2cm} \centerline{\hspace{-0.2cm}
\epsffile{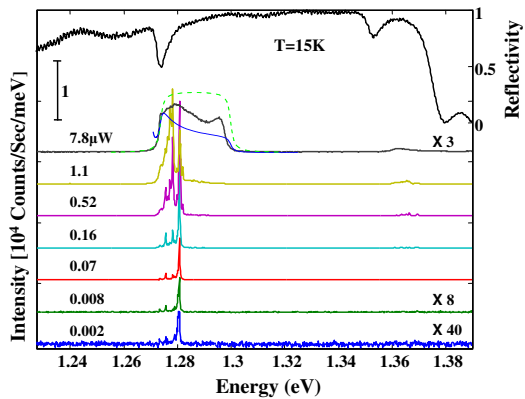}} \vspace*{-2.6cm} \caption{(color online)
Measured reflectivity (upper line) and integrated PL intensity for
various excitation powers. Calculated emission rates with detuning
by varying the QD radius (solid blue line) and its material
bandgap (dashed green line) are superimposed on the measured PL
with the highest excitation power.} \label{emission}
\end{figure}

Figure \ref{emission} shows the dependence of the QD spontaneous
emission on the excitation power. At low excitation the PL is
dominated by a sharp emission line arising from the recombination
of the 1s exciton. As the excitation power is increased, the dot
populates with more carriers and recombinations of multiexcitons
from higher collective states appear in the spectrum
\cite{Dekel98, Dekel00}. For still higher excitation powers the
emission is broadened and is extended beyond the cavity mode
linewidth (see the reflectivity curve in Fig.~\ref{emission}).
This is qualitatively confirmed by calculated emission rates that
are superimposed over the high-excitation-power PL curve (topmost
PL spectrum in Fig.~\ref{emission}). As before, the detuning of
higher dot states is represented by either varying the dot size
(solid blue line) or its material bandgap (dashed green line).

\begin{figure}[!htbp]
\epsfxsize=0.9\columnwidth
\vspace*{1.8cm} \centerline{\hspace{-0.2 cm}
\epsffile{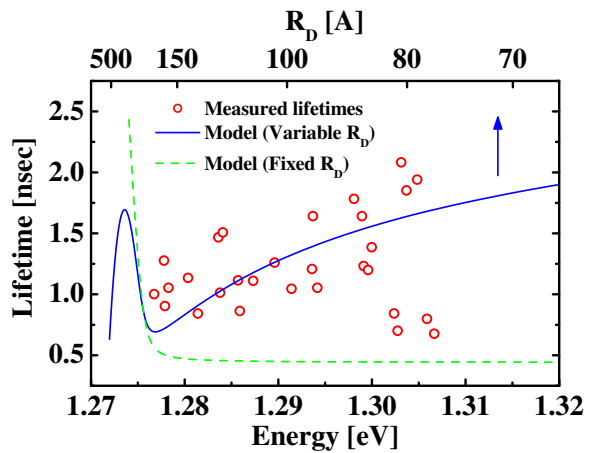}} \vspace{-2.1cm} \caption{ Exciton
radiative lifetime vs. its energy: open (red) circles are the
measured values. Solid (blue) line shows the calculation where
detuning is obtained with variable QD radius (given in the top
axis). Dashed (green) line shows the calculation with QD size
fixed to $R_{\rm D}=200\AA$ and its bandgap varied.}
\label{lifetime}
\end{figure}

In Figure \ref{lifetime} we plot the experimental and calculated
exciton lifetimes as functions of the emission spectral energy.
Each point in the figure represents a measurement from a specific
QD (open circles). The sample temperature during the measurements
was 15K. The emission lifetime was estimated from a single
exponential fit to the initial part of the measured PL decay
curves (not shown). In general, the decay curves were quite
sensitive to the excitation intensity and they were not
single-exponentials. For the measurements we used excitation
intensities which exactly saturate the PL emission. We note that
there is a considerable scatter in the measured data. We believe
that this scatter is mainly due to non-radiative processes, which
may depend on the particular environment and charge state of a
given QD. Also, as mentioned above, the measured decay times were
excitation intensity dependent. Our particular choice of
excitation intensity, for which the exciton PL saturates, may be
somewhat arbitrary. The saturation excitation may vary from one QD
to the other in a way that has little to do with the microcavity.

For the model in which the detuning is represented by varying the
QD material bandgap (dashed line), above resonance there are
always cavity modes in the 2D continuum that can couple to the
exciton due to the photon in-plane dispersion; thus the coupling
remains effective even for a large positive detuning, resulting in
a short recombination time. The emission from negatively detuned
QDs is largely inhibited consistent with the experiment where
emission from these QDs is not observed. This inhibition depends
strongly on the cavity Q-factor and would be much less pronounced
for weaker cavities. The calculations in which detunings are
represented by varying the dot size (solid line) show a different
behavior. When the QD radius is reduced (energies above resonance
in Fig.~\ref{lifetime}) the lifetime increases due to the spread
of the exciton CM wavefunction in $k_\|$ space, as explained
above. The very different lifetime behavior of the two detuning
mechanisms may aid in identifying the various QDs. The results of
both figures \ref{rate} and \ref{lifetime} seem to indicate that
the majority of QDs have similar composition and their variations
are mainly in size.

We note that the increase in radiative lifetime with reduction of
the QD size is suppressed in the limit of strong confinement. This
is because the oscillator strength ($f_\|$) increases due to the
localization of the relative motion wavefunction. This effect is
superseded by the breakdown of the selection rule caused by the
spread of the CM k-space wavefunction outside $q_{\rm x}$. Also
note that the calculated linewidths due to spontaneous emission
are at most $\gamma_{\rm SE} \lesssim 1.6 \mu{\rm eV} \ll
\gamma_c$ thus we are always in the weak coupling regime where our
approximate solution is valid.

\begin{figure}[!htbp]
\epsfxsize=0.7\columnwidth
\vspace*{0.2cm} \centerline{\hspace{-1.0cm} \epsffile{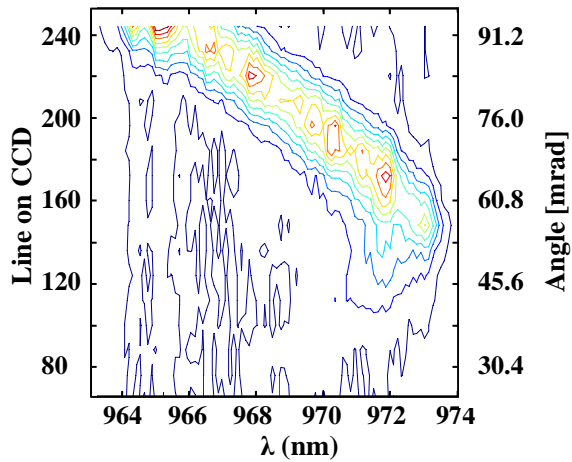}}
\vspace*{-0.2cm} \caption{(color online) Contour plot of the PL
intensity from a highly excited single resonant QD in a planar
microcavity as a function of photon wavelength and emission angle.
Each contour represents 10\% of the maximal intensity.}
\label{dudi3}
\end{figure}

\begin{figure}[!htbp]
\epsfysize=1.2\columnwidth
\vspace*{-0.2cm} \centerline{\hspace{-0.0cm}
\epsffile{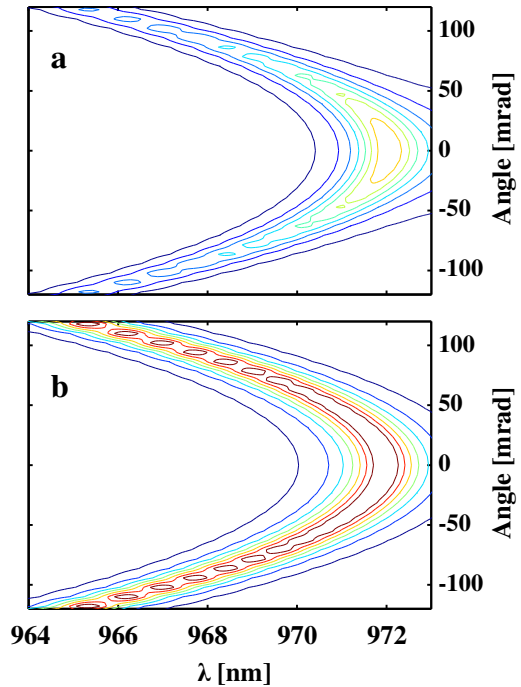}} \vspace*{-0.7 cm} \caption{Contour plot of
the calculated emission rate distributions as a function of photon
wavelength and emission angle (${\bf k}_\|$). Each contour
represents 10\% of the maximal rate. Detuning is obtained by (a)
varying the QD radius, and (b) varying the QD material bandgap and
fixing the dot radius to $R_{\rm D}=200\AA$.} \label{angular}
\end{figure}

Next, angle resolved spectroscopy was performed on a single QD
using high power excitation ($8\mu W$). In these conditions the QD
is populated with many carriers resulting in many available
recombination channels. The PL spectrum therefore reflects the
density of electromagnetic modes in the cavity. Fig.~\ref{dudi3}
shows a contour plot giving the measured PL intensity of a highly
excited single QD, as registered by the CCD array camera at the
exit of the monochromator. The horizontal axis gives the spectral
dispersion while the vertical axis gives the angular distribution.
Since the $k_z$ part of the emission is fixed by the resonance
condition to the cavity mode, this angle corresponds to the
in-plane wave vector of the emission. From the geometry and lenses
used for the imaging we estimate that each row in the CCD camera
is equivalent to 0.38mrad. The figure clearly demonstrates the
in-plane dispersion of the cavity mode. The peaks in the emission
intensity shown in the figure are most likely related to
recombinations from various carrier configurations, e.g. the
neutral and  charged excitons, biexcitons and higher order
multiexcitons ~\cite{Dekel98, Dekel00}, of the single QD. These
multiexciton lines appear due to the high excitation power used in
this experiment. Since our model accounts only for the neutral
exciton line, we cannot compare our experimental data directly
with the model calculations. Nevertheless, by taking the
derivative of Eq.~(\ref{SE2}) with respect to the ${\bf k}_\|$
space area (using $q=q_{\rm x} \sin \theta$), we can achieve a
similar in-plane dispersion of the cavity mode, as shown in
Fig.~\ref{angular}. The models of QDs in which the size and the
material bandgap were varied to represent the detuning show
similar dispersion behavior. The difference between the two cases
is mainly that in the case of fixed QD size the emission extends
over a larger energy range (Fig.~\ref{angular}b), which can be
explained using the preceding arguments. We attribute the
difference between the symmetric shape of the calculated angle
resolved spectrum and the asymmetric shape of the measured one to
the finite aperture of the confocal setup. The data was taken from
a single QD that was positively detuned with respect to the MC
resonance. Due to the relatively large spatial distribution of the
emission from the QD (see section \ref{resultsB}), we could not
collect the light from both negative and positive emission angles.

\subsection{Spatial distribution of the Emission}
\label{resultsB}

\begin{figure}[!htbp]
\epsfxsize=0.75\columnwidth
\vspace*{0.2cm} \centerline{\hspace{-0.5cm}
\epsffile{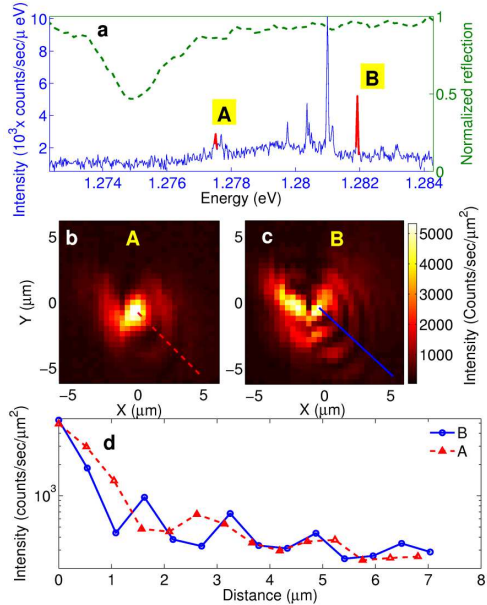}}
\caption{(color online) (a) Spatially integrated emission spectrum
measured from a single QD under moderate excitation power. The
dashed (green) line shows the normalized reflection of the sample;
Wavelength selective spatial image measured for (b) spectral line
A; and (c) spectral line B; (d) Emission intensities along the
diagonals which are marked by dashed (red) and solid (blue) lines
as functions of the distance from the image center.}
\label{image-exp}
\end{figure}

We now turn to examine the in-plane spatial spread of a single QD
emission. Fig.~\ref{image-exp}a shows a spatially integrated
emission spectrum from a particular QD under moderate excitation
power. Few discrete spectral lines are observed. Each spectral
line corresponds to a particular emission line arising from the
recombination of a ground state e-h pair from different many
carriers' collective states.\cite{Dekel98} For our purpose it
suffices to note that each line is spectrally detuned differently
from the cavity mode. In order to estimate the spectral detuning
of each line the reflectivity spectrum from this sample is
overlaid on the emission spectrum in Fig.~\ref{image-exp}a. The
least and most detuned lines are marked in the figure by A and by
B respectively. Wavelength selective spatial images of the least
(A) and most (B) detuned spectral lines are shown in
Fig.~\ref{image-exp}b and Fig.~\ref{image-exp}c, respectively. The
intensity distributions of the electromagnetic fields associated
with the emission lines are essentially symmetric, though some
obscuration, which can be partially attributed to mechanical
drift, is present. The emission patterns contain a central strong
spot surrounded by concentric rings with decreasing intensities.
The intensities along the diagonals, which are marked by dashed
(red) and solid (blue) lines in Figs.~\ref{image-exp}b and
\ref{image-exp}c, respectively, are displayed as functions of the
in-plane distance from the image center in Fig.~\ref{image-exp}d.

\begin{figure}[!htbp]
\epsfxsize=0.75\columnwidth
\vspace*{0.2cm} \centerline{\hspace{-0.5cm}
\epsffile{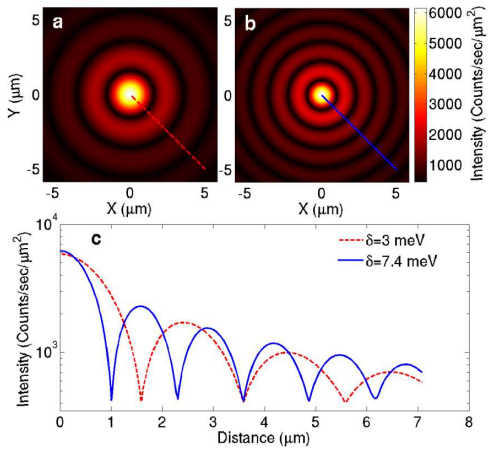}}
\caption{(color online) Calculated spatial spread of the emission
from a single QD detuned from resonance by (a)  $\delta = 3{\rm
meV}$; (b) $\delta = 7.4{\rm meV}$, corresponding to the measured
spectral lines marked A and B in Fig.~\ref{image-exp},
respectively. Detuning is modeled by varying the QD material
bandgap while the QD radius is fixed to $R_{\rm D}=200{\rm \AA}$);
(c) Calculated emission intensities as functions of the distance
from the image center. (Constant background of 400 counts was
added to the calculated emission to facilitate comparison with the
measured data).} \label{image-calc}
\end{figure}
Figs.~\ref{image-calc}a and \ref{image-calc}b show the calculated
images for the detunings of line A and line B respectively. Since
both spectral lines originate from the same QD, we used different
bandgaps to obtain the detuning. The images were generated by
calculating the in-plane Fourier transform of the integrand in
Eq.~\ref{SE}. The rates were calibrated so that their spatially
integrated emission would match the calculated values given in
Fig.~\ref{rate}. Fig.~\ref{image-calc}c shows the intensities
along the diagonals of the calculated images as functions of the
in-plane distance from the image center. In both measured and
calculated images it is evident that when the QD is close to
resonance with the cavity mode, its emission exhibits an in-plane
spatial spread that is much larger than the QD size. The closer
the emission energy to resonance is the larger is its spatial
extent, and vise versa, the larger the detuning is, the more
localized the emission pattern becomes. Calculated images, where
detuning was modeled by varying the dot size, yielded similar
behavior of the emission pattern, implying that the PL spatial
spread is mainly a cavity effect depending only on detuning. In
general, the calculated images agree with the experimental
observations, both in their spatial extent and in their dependence
on the detuning, as can be concluded by comparing
Fig.~\ref{image-calc} with Fig.~\ref{image-exp}.

\begin{figure}[!htbp]
\epsfxsize=0.75\columnwidth
\vspace*{-0.1cm} \centerline{\hspace{-0.5cm} \epsffile{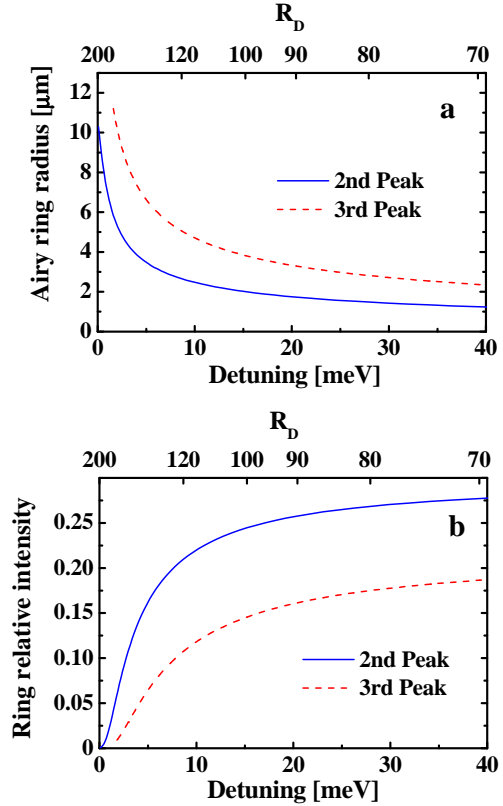}}
\caption{(color online) (a) Calculated positions of second and
third rings in the emission distribution; (b) Calculated emission
intensity of second and third rings normalized to central emission
peak.} \label{peaks}
\end{figure}

A discernible feature in Figs.~\ref{image-exp}d and
\ref{image-calc}c is the dependence of the spacing between the
rings in the PL images on the detuning. The rings become more
pronounced and denser for increasing positive detuning, where
larger $k_\|$ are needed for the emission. This gives rise to
oscillations that are superimposed on the central emission line
due to contributions arising from larger $k_\|$ modes. The
dependencies of the calculated locations of the 2nd and 3rd peaks
on detuning are shown in Figure \ref{peaks}a, and their
intensities normalized to the central peak are shown in
Fig.~\ref{peaks}b, confirming these observations. Identical
results are obtained for both detuning models, suggesting again
that this emission pattern is a cavity effect, depending only on
detuning. Together with the central peak spread, the locations of
the emission rings may aid in evaluating the QD's detuning. We
note, though, that as Fig.~\ref{peaks}b suggests, for spectral
lines that are close to resonance with the cavity mode, the
relative intensity of the rings emission to the central spot
decreases rapidly, making them difficult to measure.

\section{Summary}
\label{summary}

In this work we have studied the interaction between the
electromagnetic modes in a planar cavity and excitons in single
QDs embedded at the antinode of the cavity.

We find that the spontaneous emission outside the microcavity is
enhanced considerably for dots in resonance with the cavity mode.
The emission intensity rapidly decreases with detuning for lines
that are positively detuned from the cavity mode. Emission from
negatively detuned QDs is largely inhibited, and we were unable to
observe any spectral emission below the cavity mode energy.
Lifetime measurements have shown a decrease by a factor of roughly
two for the lifetime of resonant QDs as compared with QDs that are
20 meV detuned from the cavity line. In-plane dispersion of cavity
photons were directly observed by strong excitation of the QD,
thereby transforming it into a broadband light source. Finally, we
obtained near field images of the spatial emission distributions
from single QDs, showing a large spread over several microns in
the case of resonant QDs. These images have also shown emission
rings with separation and intensity that depend on the detuning of
the QD from the cavity mode.

All of these features were qualitatively accounted for by
calculations where detuning from the cavity mode was modeled by
either varying the QD size or its bandgap. For most of the
experimental data we have found better agreement with the first
mechanism, suggesting that most of the QDs in strain driven self
assembled samples vary in size rather than in composition. The
range of $90 {\rm \AA} \lesssim R_D\lesssim 300 {\rm \AA}$ that
accounts for the observed phenomena is consistent with typical
self assembled InGaAs/GaAs quantum dots.

\acknowledgments

This work was supported by ONR, DARPA and the ONR Nanoscale
Electronics Program. G.~R.~greatfully acknowledges the NRL/NRC
Research Associateship. The work at the Technion was supported by
the US-Israel Binational and by the Israeli Science Foundations.

\begin{widetext}

\section*{Appendix}

\renewcommand{\theequation}{A-\arabic{equation}}
\setcounter{equation}{0}

Here we provide some details of the calculation of the exciton's
relative motion energy.

\begin{eqnarray}
E_{\rm r} &\!\!=\!\!& \langle \chi_e \chi_h \phi |H_r| \phi \chi_h
\chi_e \rangle \nonumber \\
&\!\!=\!\!& 2\pi {\cal N}_r^2 (1+\sigma) \frac{\alpha^2}{\eta^{2
\alpha}} \int dr r^{2\alpha-1} {\rm e}^{-2(r/\eta)^\alpha}+ k_e^2
{\cal N}_{z_e}^2 \left[
\sigma_{ez}\frac{L}{2}+\frac{V_e^z}{\kappa_e
(k_e^2+\kappa_e^2)}\right] + k_h^2 {\cal N}_{z_h}^2 \left[
\sigma_{hz} \frac{L}{2}+\frac{V_h^z}{\kappa_h
(k_h^2+\kappa_h^2)}\right]- \nonumber \\
&& \!\! 4\pi ({\cal N}_r {\cal N}_{z_e} {\cal N}_{z_h})^2 \int dr
dq r J_0 (qr) {\rm e}^{-2(r/\eta)^\alpha} \sum_{i=1}^{4}{\cal
I}_i(q) +\frac{8\pi {\cal N}_r^2 \sigma^2}{(1+\sigma)^3 \xi^4} \!
\int \!\! dr r^3 {\rm e}^{-2(r/\eta)^\alpha}. \label{Eexp}
\end{eqnarray}
${\cal I}_i(q)$ are given by
\begin{subequations}
\begin{eqnarray}
{\cal I}_1(q)&=&\frac{L}{2q}+\frac{1}{4q^2} \left[ \sum_{i=e,h}
\frac{g_i}{k_i^2}+ \sum_{i\neq j} \frac{g_i}{k_i^2-k_j^2} \left(
\frac{f_j}{q^2+4k_e^2}-1 \right) \right] - \nonumber \\
&& {\rm e}^{-qL/2} (f_e-g_e) \frac{ \cosh(qL/2) g_h+ \sinh(qL/2)
f_h }{q^2(q^2+4k_e^2)(q^2+4k_h^2)} \\
{\cal I}_2(q)&=& \frac{2k_e^2 {\rm e}^{-qL/2}}{k_e^2+\kappa_e^2}
\frac{\sinh(qL/2) f_h +\cosh(qL/2)g_h}{q(q+2\kappa_e)(q^2+4k_h^2)}
\\
{\cal I}_3(q)&=&\frac{2k_h^2 {\rm e}^{-qL/2}}{k_h^2+\kappa_h^2}
\frac{\sinh(qL/2) f_e + \cosh(qL/2)g_e}{q(q+2\kappa_h)(q^2+4k_e^2)} \\
{\cal I}_4(q)&=& \frac{4 k_e^2 k_h^2}{(k_e^2+\kappa_e^2)
(k_h^2+\kappa_h^2)}
\frac{\kappa_e+\kappa_h+q}{(\kappa_e+\kappa_h)(2\kappa_e+q)(2\kappa_h+q)}
\end{eqnarray}
\label{I}
\end{subequations}
where
\begin{equation}
f_i=4k_i^2+q^2+q^2 \cos(k_i L) \ \ , \ \ g_i=2q k_i \sin(k_i L).
\end{equation}

\end{widetext}



\begin{references}

\bibitem{Michler-Santori-Regelman} P.~Michler, A.~Imamo\u{g}lu,
M.~D.~Mason, P.~J.~Carson, G.~F.~Strouse, and S.~K.~Buratto,
Nature (London) {\bf 406}, 968 (2000); C.~Santori, M.~Pelton,
G.~Solomon, Y.~Dale, and Y.~Yamamoto, Phys.~Rev.~Lett.~{\bf 86},
1502 (2001); D.~V.~Regelman, U.~Mizrahi, D.~Gershoni,
E.~Ehrenfreund, W.~V.~Schoenfeld, and P.~M.~Petroff, {\it ibid}
{\bf 87}, 257401 (2001).
%
\bibitem{Kiraz} A.~Kiraz, M.~Atat\"{u}re, and A.~Imamoglu,
Phys.~Rev.~A {\bf 69}, 032305 (2004).
%
\bibitem{Knill} E.~Knill, R.~Laflamme, and G.~J.~Milburn,
Nature (London) {\bf 409}, 46 (2001).
%
\bibitem{Santori} C.~Santori, D.~Fattal, J.~Vu\u{c}kovi\'{c},
G.~S.~Solomon, and Y.~Yamamoto, Nature (London) {\bf 419}, 594
(2002).
%
\bibitem{Vahala} K.~J.~Vahala, Nature (London) {\bf 424}, 839
(2003).
%
\bibitem{Gerard} J.~M.~G\'{e}rard, B.~Sermage, B.~Gayral,
B.~Legrand, E.~Costard, and V.~Thierry-Mieg, Phys.~Rev.~Lett.~{\bf
81}, 1110 (1998).
%
\bibitem{Pelton} M.~Pelton, C.~Santori, J.~Vu\u{c}kovi\'{c},
B.~Zhang, G.~S.~Solomon, J.~Plant, and Y.~Yamamoto,
Phys.~Rev.~Lett.~{\bf 89}, 233602 (2002).
%
\bibitem{Garyal} B.~Gayral, J.~M.~G\'{e}rard, A.~Lema\t{it}re,
C.~Dupuis, L.~Manin, and J.~L.~Pelouard, Appl.~Phys.~Lett.~{\bf
75}, 1908 (1999).
%
\bibitem{Yoshie} T.~Yoshie, A.~Scherer, J.~Hendrickson,
G.~Khitrova, H.~M.~Gibbs, G.~Rupper, C.~Ell, O.~B.~Shchekin, and
G.~Deppe, Nature (London) {\bf 432}, 200 (2004).
%
\bibitem{Reithmaier} J.~P.~Reithmaier, G.~S\c{e}k, A.~L\"{o}ffler,
C.~Hofmann, S.~Kuhn, S.~Reitzenstein, L.~V.~Keldysh,
V.~D.~Kulakovskii, T.~L.~Reinecke, and A.~Forchel, Nature (London)
{\bf 432}, 197 (2004).
%
\bibitem{Badolato} A. Badolato, K. Hennessy, M. Atatu(re,
J. Dreiser, E. Hu, P. M. Petroff, A. Imamog"lu, Science {\bf 308},
1138 (2005).
%
\bibitem{Andreani} L.~C.~Andreani, G.~Panzarini, and
J-M.~G\'{e}rard, Phys.~Rev.~B {\bf 60}, 13276 (1999).
%
\bibitem{Bjork91} G.~Bj\"{o}rk, S.~Machida, Y.~Yamamoto, and K.~Igeta,
Phys.~Rev.~A {\bf 44}, 669 (1991).
%
\bibitem{Bennett} A.~J.~Bennett, D.~C.~Unitt, P.~See,
A.~J.~Shields, P.~Atkinson, K.~Cooper, and D.~A.~Ritchie,
Appl.~Phys.~Lett.~{\bf 86}, 181102 (2005).
%
\bibitem{Weis} C.~Weisbuch, M.~Nishioka, A.~Ishikawa, Y.~Arakawa,
Phys.~Rev.~Lett, {\bf 69}, 3314 (1992)
%
\bibitem{Savona1} V.~Savona, Z.~Hradil, A.~Quattropani, and
P.~Schwendimann, Phys.~Rev.~B {\bf 49}, 8774 (1994).
%
\bibitem{Savona2} V.~Savona, F.~Tassone, C.~Piermarocchi, A.~Quattropani, and
P.~Schwendimann, Phys.~Rev.~B {\bf 53}, 13051 (1996).
%
\bibitem{Citrin} D.~S.~Citrin IEEE J.~Quant.~Elec.~{\bf 30}, 997
(1994).
%
\bibitem{Sug2} M.~Sugawara,Jpn.~J.~Appl.~Phys.~{\bf 36}, 2151
(1997).
%
\bibitem{Andrews} J.~T.~Andrews, P.~Sen, and R.~R.~Puri
J.~Phys.~Cond.~Matt. {\bf 11}, 6287 (1999).
%
\bibitem{Hu} C.~Y.~Hu, H.~Z.~Zheng, J.~D.~Zhang, H.~Zhang,
F.~H.~Yang, and Y.~P.~Zeng, Appl.~Phys.~Lett.~{\bf 82}, 665
(2003).
%
\bibitem{Garcia} J.~M.~Garcia, T.~Mankad, P.~O.~Holtz, P.~J.~Wellman, and
P.~M.~Petroff, Appl.~Phys.~Lett.~{\bf 72}, 3172 (1998).
%
\bibitem{Que} W.~Que, Phys.~Rev.~B {\bf 45}, 11036 (1992).
%
\bibitem{CM} For a general lateral poetential, seperation of the CM- and relative-motions is
valid only when the QD radius is much larger than the exciton Bohr
radius.
%
\bibitem{Sug1} M.~Sugawara, Phys.~Rev.~B {\bf 51}, 10743 (1995).
%
\bibitem{GreBaj} R.~L.~Greene, K.~K.~Bajaj and D.~E.~Phelps,
Phys.~Rev.~B {\bf 29}, 1807 (1984).
%
\bibitem{GruBim} M.~Grundmann and D.~Bimberg, Phys.~Rev.~B {\bf
38}, R13486 (1988).
%
\bibitem{GershoniJQE} D.~Gershoni, C.~H.~Henry, and G.~A.~Baraff, IEEE journal of quantum electronics {\bf 29}, 2433 (1993).
%
\bibitem{Jackson} J.~D.~Jackson, {\em Classical Electrodynamics}, 2nd edition
(John Wiley, New York, 1975).
%
\bibitem{Andreani90} L.~C.~Andreani and A.~Pasquarello,
Phys.~Rev.~B{\bf 42}, 8928 (1990).
%
\bibitem{Dekel00} E.~Dekel, D.~Gershoni, E.~Ehrenfreund, J.~M.~Garcia,
and P.~M.~Petroff, Phys.~Rev.~B ~{\bf 61}, 11009 (2000).
%
\bibitem{Dekel98} E.~Dekel, D.~Gershoni, E.~Ehrenfreund, D.~Spektor, J.~M.~Garcia,
and P.~M.~Petroff, Phys.~Rev.~Lett.~{\bf 80}, 4991 (1998).
%

\end{references}
\end{document}